\documentclass[acmsmall]{acmart} 
\AtBeginDocument{%
  }

\setcopyright{acmlicensed}
\copyrightyear{2018}
\acmYear{2018}
\acmDOI{XXXXXXX.XXXXXXX}
\acmConference[Conference acronym 'XX]{Make sure to enter the correct
  conference title from your rights confirmation email}{June 03--05,
  2018}{Woodstock, NY}
\acmISBN{978-1-4503-XXXX-X/2018/06}

\newcommand\sys{\textit{MCPacific{}}}
\usepackage{xcolor}
\usepackage{tabularx}

\usepackage{float}
\usepackage{algorithm}
\usepackage{algpseudocode}
\usepackage{caption}

\floatstyle{ruled}
\restylefloat{algorithm}

\algrenewcommand\algorithmicrequire{\textbf{Input:}}
\algrenewcommand\algorithmicensure{\textbf{Output:}}

\usepackage{hyperref}
\usepackage{url}
\usepackage[T1]{fontenc}
\usepackage{amsmath,mathtools}
\usepackage{graphicx,booktabs,tabularx,array}
\usepackage{wrapfig}
\usepackage{microtype}
\usepackage{subcaption}
\usepackage{booktabs}
\usepackage{tabularx}
\usepackage{makecell}
\usepackage{threeparttable}
\usepackage{pifont}
\usepackage{booktabs,tabularx,array,makecell,xcolor}

\usepackage[most]{tcolorbox}
\usepackage{multirow}

\newcolumntype{Y}{>{\raggedright\arraybackslash}X}

\newcommand{\figplaceholder}[3]{%
  \IfFileExists{#1}{\includegraphics[width=\linewidth,height=#2,keepaspectratio]{#1}}{%
    \fbox{\parbox[c][#2][c]{\dimexpr\linewidth-2\fboxsep-2\fboxrule\relax}{%
    \centering\small Figure placeholder\\[4pt]#3}}}}
\newif\ifdraftnote

\definecolor{MCPTeal}{HTML}{087F8C}
\definecolor{MCPAmber}{HTML}{9A6700}
\definecolor{MCPGray}{HTML}{6B7280}

\definecolor{FindingBg}{HTML}{F7F9FC}
\definecolor{FindingBar}{HTML}{A9BAD8}
\newcounter{finding}

\newcommand{\mcpNo}{%
\textcolor{red}{\ding{55}}}
\newcommand{\mcpYes}{%
  \textcolor{MCPTeal}{\ensuremath{\checkmark}}}
\newcommand{\mcpPartial}{%
  \textcolor{MCPAmber}{\ensuremath{\triangle}}}
\newcommand{\mcpDash}{%
  \textcolor{MCPGray}{\textemdash}}

\newcommand{\mcpHead}[1]{%
  {\scriptsize\bfseries\makecell[c]{#1}}}

\newcommand{\mcpEntryC}[2]{%
  \textbf{#1}\hspace{2pt}{\tiny\citep{#2}}}

\providecommand{\answerbox}[2]{%
  \par\smallskip\noindent
  \fcolorbox{MCPGray}{gray!8}{%
    \parbox{\dimexpr\linewidth-2\fboxsep-2\fboxrule\relax}{%
      \textbf{Answer to #1:} #2}}%
  \par\smallskip}

\begin{document}

\title{Understanding the Hierarchical Structure and Functional Landscape of the Model Context Protocol Ecosystem}

\author{Tingxuan Tang}
\affiliation{%
  \institution{William \& Mary}
  \city{Williamsburg}
  \state{Virginia}
  \country{USA}
}

\author{Zilong Chen}
\affiliation{%
}

\author{Yue (Luna) Xiao}
\affiliation{%
  \institution{William \& Mary}
  \city{Williamsburg}
  \state{Virginia}
  \country{USA}
}

\renewcommand{\shortauthors}{Tang et al.}

\begin{abstract}
AI agents increasingly rely on tools exposed through the Model Context Protocol (MCP) to complete user tasks. Hundreds of thousands of MCP servers are listed across marketplaces, yet they are organized only by coarse, marketplace-specific server categories. This makes it difficult for agents and users to identify tools for a given operation, find functional alternatives, and assess how those alternatives differ.
We present \sys{}, the largest tool-level, cross-marketplace map of the MCP ecosystem. 
\sys{} collects 368{,}754 MCP server listings corresponding to 124{,}267 unique servers across 17 marketplaces, statically extracts 1{,}328{,}233 tool specifications from these servers in seven languages, and organizes them into a hierarchical functional taxonomy of 58{,}915 capabilities.
We construct the taxonomy through an iterative LLM-driven design--test--refine process and map the full corpus to it using calibrated embedding routing.
Our study reveals that MCP extends well beyond developer tooling, with 85\% of tools serving other domains. Functional alternatives are widespread but unevenly distributed: 98.5\% of tools have at least one alternative, yet nearly a quarter of capabilities are supported by only one tool. Functionally comparable tools also differ in security alerts, code complexity, and project maintenance, with complexity differing by more than 2.5$\times$ in 41\% of comparable tool pairs. Finally, presenting candidate tools through the taxonomy rather than a flat list improves task completion rate across all four evaluated models, with gains of up to 12 percentage points in Pass@0.75 for crowded candidate sets.
%

\end{abstract}

\ccsdesc[500]{Model Context Protocol}
\ccsdesc[300]{Hierarchical Taxonomy}
\ccsdesc[300]{Tool Alternatives}
\ccsdesc{Benchmark}

\keywords{taxonomy construction, multi-agent systems, large language models}

\received{20 February 2007}
\received[revised]{12 March 2009}
\received[accepted]{5 June 2009}

\maketitle

\section{Introduction}
\label{sec:intro}

Large language model (LLM) agents increasingly act on the world through external tools: they search the web, query databases, edit files, and operate software services~\citep{3666122.3669119,yao2022react}.
The Model Context Protocol (MCP)~\citep{anthropic2024mcp} has quickly become a widely adopted standard for exposing such tools: an MCP server publishes a set of tools, each with a name, a natural-language description, and an input schema, and any MCP-compatible agent can discover and invoke them~\citep{mcp2026spec}.
Within two years, this standard has produced an open ecosystem of more than 100{,}000 independently developed servers, advertised across many marketplaces.

This openness turns tool use into a \emph{search} problem.
Before an agent can call a tool, it must find one that provides the needed capability; and when that tool is unavailable, deprecated, or insecure, it must find another that does the same job.
Consider an agent tasked with sending an email through Gmail. In our taxonomy, 18 tools from different MCP servers provide this same capability, such as \texttt{send\_gmail\_message}, \texttt{send\_via\_gmail}, \texttt{gmail\_send}, and \texttt{gmail\_send\_email}.
The corresponding repositories range from 0 to 1{,}015 stars and were last updated between 15 and 480 days before collection. 
They are maintained by either organizations (e.g., Cloudgentic AI) or individual developers and implemented in diverse languages, including Python, Go, TypeScript, and C\#.
Despite relying on different providers and input schemas, they perform the same core function and may serve as alternatives. However, such relationships are not visible in existing marketplaces, which organize servers under coarse, marketplace-specific services, such as Communication and Workplace \& Productivity, rather than indexing the capabilities of individual tools which are often documented only in source code.
As a result, neither agents nor users can answer basic questions: \textit{Which capabilities does the ecosystem offer? Which tools can substitute for one another? And how do such alternatives differ?}

%

Existing work does not answer these questions.
Agent benchmarks~\citep{wang2026mcp,luo2025mcp,bandi2026mcp,mo2026livemcpbench} evaluate tool use within a limited set of curated servers, while tool-retrieval studies~\citep{qin2024toolllm,fei2025mcp,wang2026mcpflow} rank tools within fixed collections containing at most 177{,}436 tools, roughly 13\% of the tools we extract, and organize them only into high-level task domains rather than tool-level functional categories.
Ecosystem studies~\citep{guo2025measurement,lin2025mcpcorpus,chen2026rethinking,toeppe2026mcpimplementations,10.1145/3796519} characterize servers through marketplace metadata, deployment properties, or security analyses, but remain largely at the server level, even though a single server may bundle dozens of tools that perform distinct operations.
What is missing is a \emph{tool-level, functional} view of the ecosystem: one that tells which operations exist and groups tools that perform the same operation, regardless of which server or marketplace lists them.

Building such a view poses three challenges.
\textbf{(C1) Fragmented sources.} Servers are listed across heterogeneous marketplaces with different interfaces, duplicate entries, and inconsistent metadata.
\textbf{(C2) Hidden tools.} Marketplaces rarely list the tools a server exposes; recovering them requires analyzing source code written in many languages, SDKs, and registration styles.
\textbf{(C3) No functional organization.} Grouping more than a million tools by what they do requires categories that are fine-grained enough to separate operations, consistent across domains, and tested against real tools, at a cost that scales.
We address these challenges with \sys{}, a tool-level, cross-marketplace map of the MCP ecosystem (Figure~\ref{fig:overview}).
\sys{} collects 368{,}754 listings from 17 marketplaces and resolves them to 124{,}267 unique source-available projects (C1).
It then statically analyzes their source code with 30 language-specific recognition patterns, extracting 1{,}328{,}233 tool specifications from 110{,}482 projects in seven languages (C2), 7.5$\times$ more tools than the largest prior collection~\citep{stein2026ai}.
Finally, it organizes the tools into a hierarchical functional taxonomy of 58{,}915 leaf capabilities (C3).
Inspired by recent work on software taxonomies~\citep{lu2026atlas}, we build the taxonomy top-down: at each node, an LLM designs categories from sampled tools, tests them on a separate pilot sample, and refines them until they cover the tools with clear boundaries. We then map the full corpus with calibrated embedding routing and resolve the remaining tools through human review.
Each leaf groups \emph{functionally comparable} tools, i.e., tools that perform the same operation on the same kind of resource, even if they differ in provider, schema, or implementation.
An evaluation on a uniform sample shows that 90.19\% of eligible tools reach a leaf, 87.00\% are assigned to the correct leaf, and 85.00\% of same-leaf pairs are functionally comparable.

Using \sys{} as a measurement instrument, we conduct the largest empirical study of MCP tool functionality to date, organized around four research questions and their key findings:

\vspace{2pt}
\noindent\textbf{RQ1 (Functional landscape):} \textit{``Which capabilities does the ecosystem offer, and how are tools distributed across them?''} Our study shows that 85\% of tools serve domains beyond software development, including finance, media, and healthcare. At the same time, large domains fragment into thousands of narrow capabilities that coarse marketplace categories cannot distinguish.

\vspace{2pt}
\noindent\textbf{RQ2 (Functional comparability):} \textit{``How often do tools have functionally comparable alternatives, and which capabilities lack them?''} We find that functionally comparable alternatives are widespread but unevenly distributed: 98.5\% of tools have at least one alternative and 74.1\% have at least 20, yet nearly one quarter of capabilities are represented by only a single tool.

\vspace{2pt}
\noindent\textbf{RQ3 (Alternative heterogeneity):} \textit{``How do functionally comparable tools differ in security-relevant properties, code complexity, and project characteristics?''} We find that comparable tools can differ substantially: security alerts often concentrate in only a subset of alternatives, cyclomatic complexity differs by more than 2.5$\times$ for 41\% of tool pairs, and in 60\% of the capabilities we examine, some candidate projects are actively maintained while others are not.

\vspace{2pt}
\noindent\textbf{RQ4 (Downstream task utility):} \textit{``Does organizing candidate tools by the taxonomy help agents complete tasks?''} We find that taxonomy-guided organization improves task completion on MCP-Atlas~\citep{bandi2026mcp} for all four evaluated models, with gains of up to 12 points in Pass@0.75 as the candidate set becomes more crowded.

\vspace{4pt}\noindent\textbf{Contributions.} This paper makes the following contributions:

\vspace{2pt}
\noindent\textbullet\ \textit{Dataset.} We build the largest tool-level dataset of the MCP ecosystem, comprising 1{,}328{,}233 tool specifications statically extracted from 124{,}267 unique MCP servers listed across 17 marketplaces.

\vspace{2pt}
\noindent\textbullet\ \textit{Taxonomy.} We construct a hierarchical functional taxonomy of MCP tools containing 58{,}915 capabilities, together with a scalable and validated construction pipeline that combines LLM-assisted category design and evaluation with calibrated embedding-based routing.

\vspace{2pt}
\noindent\textbullet\ \textit{Empirical study.} We conduct the largest characterization to date of MCP tool functionality, functional comparability, and heterogeneity among alternatives, revealing implications for tool discovery, selection, and ecosystem governance.

\vspace{2pt}
\noindent\textbullet\ \textit{Downstream utility.} We demonstrate that taxonomy-guided tool organization improves agent task completion across all evaluated models, particularly as the candidate tool space grows.

\section{Background and Related Work}
\label{sec:related}

In this section, we first introduce the MCP ecosystem and then review prior work in two closely related areas: (1) MCP datasets and benchmarks and (2) software taxonomy construction.

\subsection{Background.}

\noindent\textbf{MCP protocol, servers, and tools.}
The Model Context Protocol (MCP), introduced by Anthropic in November 2024, is an open standard for connecting LLM-based applications to external data and functionality~\cite{anthropic2024mcp}.
An MCP \emph{host} uses \emph{clients} to communicate with \emph{servers} that expose capabilities from external software and services, such as web browsers, IDEs, and databases~\cite{mcp2026architecture}.
A server can expose multiple tools. Each tool has a name and JSON input schema and may also provide a natural-language description and output schema~\cite{mcp2026tools,mcp2026schema}.

\noindent\textbf{Tool discovery and selection.}
In a typical agent workflow, the host retrieves available tool specifications through \texttt{tools/list} and provides them to the LLM.
Given a user request and context, the LLM uses the tool metadata to select a tool and generate its arguments.
The client invokes the tool through \texttt{tools/call}, and its result is returned to the LLM for further reasoning or actions~\cite{mcp2026architecture,mcp2026tools}.

\noindent\textbf{MCP marketplaces.}
MCP marketplaces, such as Glama~\cite{glamaMcpFaq} and Cline's MCP Marketplace~\cite{clineMcpMarketplace}, are discovery platforms operated by companies, communities, or independent maintainers.
Server developers can also submit servers, while marketplace operators index listings with descriptions, repository links, and platform-specific categories~\cite{clineMcpMarketplace,glamaMcpFaq}.
Users browse or search these catalogs to find servers to install or connect to their agents.
However, marketplace organization typically relies on coarse server-level categories, which do not capture fine-grained tool functionality or functional comparability across servers~\cite{glamaMcpFaq}. To address this gap, our work constructs a tool-level functional taxonomy to support fine-grained tool discovery and comparison across MCP servers.

\subsection{MCP Datasets and Benchmarks}

Existing MCP datasets and benchmarks primarily support server-level characterization or task-oriented tool use, as summarized in Table~\ref{tab:mcp-toolscape-comparison}.

\noindent\textbf{Ecosystem datasets and measurement.}
MCPCorpus~\cite{lin2025mcpcorpus} consolidates server and client metadata, while Toeppe et al.~\cite{toeppe2026mcpimplementations} construct a repository dataset with implementation evidence and operational-role labels.
Measurement studies examine marketplace growth and overlap~\cite{guo2025measurement}, security and maintainability~\cite{10.1145/3814959}, lifecycle threats~\cite{10.1145/3796519}, and registry and server vulnerabilities~\cite{li2026mcpinspect}.
MCPZoo~\cite{chen2026rethinking} additionally provides runnable server deployments for security measurement and scanner assessment.
Stein~\cite{stein2026ai} analyzes 177,436 tools using occupational-task and capability classifications, focusing on high-level task domains because finer-grained occupational assignments are less reliable.
These works primarily characterize MCP at the server level, leaving tool-level measurement and functional characterization underexplored.

\noindent\textbf{MCP tool-use and retrieval benchmarks.}
MCP-Universe~\cite{luo2025mcp}, MCP-Bench~\cite{wang2026mcp}, and MCP-Atlas~\cite{bandi2026mcp} evaluate multi-step tasks using real MCP servers.
MCPToolBench++~\cite{fan2025mcptoolbench++} automates task construction, while LiveMCPBench~\cite{mo2026livemcpbench} emphasizes tool retrieval and composition.
Larger resources include MCP-Zero~\cite{fei2025mcp}, which evaluates on-demand retrieval over 2,797 tools, and MCP-Flow~\cite{wang2026mcpflow}, which collects 11,536 tools to generate training and evaluation data.
JSPLIT~\cite{antonioni2025jsplit} uses a hierarchical server taxonomy to reduce retrieval and prompt overhead, but its evaluation excludes distractors from the target server's leaf category.
These works focus on tool use for specific tasks within selected servers. In contrast, \sys{} combines automated collection across 17 marketplaces with extraction of 1,328,233 tool specifications from source code (the \textbf{largest} data collection in Table~\ref{tab:mcp-toolscape-comparison}) to characterize functional coverage, redundancy, and comparable alternatives across the ecosystem.

\subsection{Software Taxonomy Construction}

Software taxonomy construction organizes software artifacts into categories based on shared characteristics, supporting systematic understanding, discovery, and comparison~\citep{usman2017taxonomies}.
Automated approaches categorize software using semantic analysis, topic modeling, or clustering~\citep{kawaguchi2004mudablue,tian2009lact,altarawy2018lascad}, while other work organizes software domain labels through knowledge bases and human or LLM assistance~\citep{sas2023gitranking, sas2024automatic}.
More recently, ATLAS~\citep{lu2026atlas} constructs hierarchical software repository taxonomies through iterative category design, classification, and refinement, supporting repository retrieval and alternative discovery.
\sys{} adapts the iterative taxonomy construction to enable functional characterization and comparison of callable operations across MCP servers.


\begin{table}[t]
\centering
\caption{Comparison of MCP datasets and benchmarks.}
\label{tab:mcp-toolscape-comparison}

\scriptsize
\setlength{\tabcolsep}{2.1pt}
\renewcommand{\arraystretch}{1.08}
\renewcommand{\tabularxcolumn}[1]{m{#1}}

\begin{tabularx}{\linewidth}{
  @{}>{\raggedright\arraybackslash}X
  >{\centering\arraybackslash}c
  >{\centering\arraybackslash}c
  >{\centering\arraybackslash}c
  >{\centering\arraybackslash}c
  >{\centering\arraybackslash}c
  >{\centering\arraybackslash}c
  >{\centering\arraybackslash}c@{}
}
\toprule

\textbf{Dataset}
& \mcpHead{\#Tools}
& \mcpHead{\#MCP\\servers}
& \mcpHead{Automatic\\collection}
& \mcpHead{Validated\\functional\\taxonomy}
& \mcpHead{Comparable-tool\\identification}
& \mcpHead{Interactive\\exploration}
& \mcpHead{Public\\release} \\

\midrule

\mcpEntryC{MCP-Universe}{luo2025mcp}
& 133
& 11
& \mcpNo
& \mcpNo
& \mcpNo
& \mcpYes
& \mcpYes \\

\mcpEntryC{MCPToolBench++}{fan2025mcptoolbench++}
& 87
& 12
& \mcpYes
& \mcpNo
& \mcpYes
& \mcpPartial
& \mcpYes \\

\mcpEntryC{MCP-Bench}{wang2026mcp}
& 250
& 28
& \mcpNo
& \mcpNo
& \mcpNo
& \mcpNo
& \mcpYes \\

\mcpEntryC{MCP-Atlas}{bandi2026mcp}
& 220
& 36
& \mcpNo
& \mcpNo
& \mcpNo
& \mcpNo
& \mcpYes \\

\mcpEntryC{LiveMCPBench}{mo2026livemcpbench}
& 527
& 70
& \mcpPartial
& \mcpYes
& \mcpNo
& \mcpNo
& \mcpYes \\

\mcpEntryC{MCP-Zero}{fei2025mcp}
& 2,797
& 308
& \mcpNo
& \mcpNo
& \mcpNo
& \mcpNo
& \mcpYes \\

\mcpEntryC{MCP-Flow}{wang2026mcpflow}
& 11,536
& 1,166
& \mcpYes
& \mcpNo
& \mcpPartial
& \mcpNo
& \mcpYes \\

\mcpEntryC{GitHub MCP implementations}{toeppe2026mcpimplementations}
& \mcpDash
& 1,962
& \mcpYes
& \mcpNo
& \mcpNo
& \mcpNo
& \mcpYes \\

\mcpEntryC{JSPLIT}{antonioni2025jsplit}
& \mcpDash
& $\sim$2,000
& \mcpNo
& \mcpPartial
& \mcpNo
& \mcpNo
& \mcpDash \\

\mcpEntryC{MCP ecosystem measurement}{guo2025measurement}
& \mcpDash
& 8,060
& \mcpYes
& \mcpNo
& \mcpNo
& \mcpNo
& \mcpYes \\

\mcpEntryC{MCPCorpus}{lin2025mcpcorpus}
& \mcpDash
& 13,875
& \mcpYes
& \mcpNo
& \mcpNo
& \mcpYes
& \mcpYes \\

\mcpEntryC{Public agent tools}{stein2026ai}
& 177,436 
& 19,388
& \mcpYes 
& \mcpYes 
& \mcpNo 
& \mcpPartial 
& \mcpPartial \\

\mcpEntryC{MCPInspect}{li2026mcpinspect}
& \mcpDash
& 67,057
& \mcpYes
& \mcpNo
& \mcpNo
& \mcpNo
& \mcpYes \\

\mcpEntryC{MCP Landscape}{10.1145/3796519}
& \mcpDash
& 105,378
& \mcpNo
& \mcpNo
& \mcpNo
& \mcpNo
& \mcpYes \\

\mcpEntryC{MCPZoo}{chen2026rethinking}
& \mcpDash
& 113,927
& \mcpYes
& \mcpYes
& \mcpNo
& \mcpYes
& \mcpPartial \\

\midrule

\textbf{MCPacific (ours)}
& 1,328,233 
& 368{,}754
& \mcpYes
& \mcpYes
& \mcpYes
& \mcpYes
& \mcpYes \\

\bottomrule
\end{tabularx}
\end{table}

\section{Methodology}
\label{sec:dataset}
\begin{figure*}[t]
  \centering
  \includegraphics[width=\linewidth]{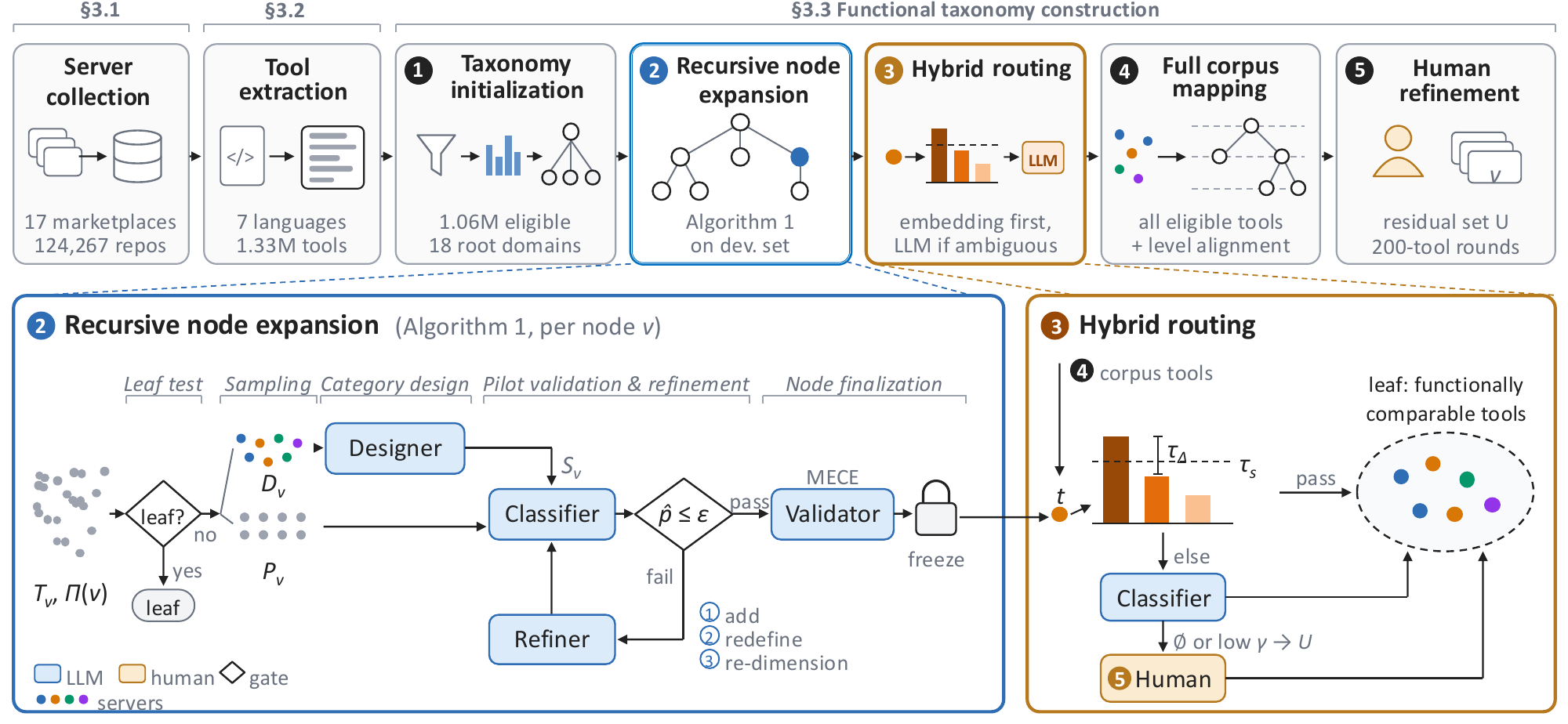}
  \caption{Overview of \sys{}. MCP servers are collected from 17
  marketplaces (\S\ref{sec:collection}), and tool specifications are
  extracted from their source code (\S\ref{sec:extraction}). The functional
  taxonomy (\S\ref{sec:taxonomy}) is initialized with 18 domains
  (\ding{202}), grown node by node on a development set (\ding{203},
  Algorithm~\ref{alg:expand-node}), and populated through hybrid routing
  (\ding{204}) and full corpus mapping (\ding{205}); human refinement
  resolves the remaining tools (\ding{206}).}
  \label{fig:overview}
  \vspace{-10pt}
\end{figure*}

Figure~\ref{fig:overview} gives an overview of \sys{}, which proceeds in three phases. First, we collect MCP servers from 17 marketplaces, deduplicate them by repository, and retrieve their source code (\S\ref{sec:collection}). Second, we statically analyze the source code to extract tool specifications in a unified five-field schema (\S\ref{sec:extraction}). Third, we organize the extracted tools into a hierarchical functional taxonomy and assign every tool to it (\S\ref{sec:taxonomy}). Each phase builds on the output of the previous one, and the resulting taxonomy serves as the measurement instrument for the analyses in Section ~\ref{sec:measurement}.

\subsection{MCP Server Collection}
\label{sec:collection}

To achieve broad coverage of publicly available MCP servers, we construct an initial pool of marketplaces and directories from prior MCP ecosystem studies~\citep{guo2025measurement,chen2026rethinking,lin2025mcpcorpus}, supplemented by community-maintained resource collections and public discussions.
We manually review these sources and prioritize diversity in platform operators, geographic regions, and functional categories, while requiring publicly accessible server listings and repository information. After consolidating duplicate or overlapping sources, we retain 17 distinct MCP marketplaces.

\setlength{\intextsep}{2pt}

\begin{wraptable}{r}{0.50\textwidth}
    \centering

    \setlength{\abovecaptionskip}{0pt}
    \setlength{\belowcaptionskip}{3pt}

    \caption{Scale of MCP server collection across multiple sources and regions.}
    \label{tab:mcp_collection}

    \scriptsize
    \setlength{\tabcolsep}{2.5pt}
    \renewcommand{\arraystretch}{1.03}

    \begin{tabularx}{\linewidth}{
        @{}
        >{\raggedright\arraybackslash}p{0.22\linewidth}
        >{\raggedright\arraybackslash}p{0.14\linewidth}
        >{\raggedright\arraybackslash}X
        >{\raggedleft\arraybackslash}p{0.13\linewidth}
        >{\raggedleft\arraybackslash}p{0.14\linewidth}
        @{}
    }
        \toprule
        \textbf{Source}
        & \textbf{Region}
        & \textbf{Operator}
        & \textbf{Raw}
        & \textbf{Valid} \\
        \midrule

        Glama
        & US
        & Glama LLC
        & 61,213
        & 57,696 \\

        Pulse MCP
        & US
        & Antanavicius et al.
        & 22,021
        & 21,355 \\

        Smithery
        & US
        & Clavia, Inc.
        & 10,338
        & 2,906 \\

        Cline MCP
        & US
        & Cline Bot Inc.
        & 199
        & 194 \\

        MCP Market
        & Canada
        & Sitka Labs
        & 30,136
        & 28,409 \\

        MCPVault
        & Sweden
        & Henrik Åberg
        & 18,114
        & 17,980 \\

        MCPNest
        & Portugal
        & Ricardo Rodrigues
        & 9,451
        & 7,563 \\

        AgentForge
        & Czech.
        & KOWEX Co.
        & 15,049
        & 57 \\

        mcpmarket.cn
        & China
        & Lingxi Yidong
        & 137,327
        & 134,905 \\

        MCP World
        & China
        & Baidu
        & 45,340
        & 44,611 \\

        SCNet MCP
        & China
        & Sugon IC
        & 13,635
        & 10,045 \\

        AIBase MCP
        & China
        & Xiamen Xianglian 
        & 10,049
        & 9,103 \\

        ModelScope
        & China
        & ModelScope
        & 9,421
        & 6,400 \\

        Tencent MCP
        & China
        & Tencent Cloud
        & 1,088
        & 971 \\

        MCP.so
        & -
        & idoubi
        & 17,858
        & 15,922 \\

        MCPServers.org
        & -
        & wong2
        & 11,811
        & 9,060 \\

        AiMCP
        & -
        & Hekmon
        & 1,683
        & 1,577 \\

        \midrule

        \textbf{Total}
        & \textemdash
        & \textemdash
        & \textbf{414,733}
        & \textbf{368,754} \\

        \textbf{Unique}
        & \textemdash
        & \textemdash
        & \textemdash
        & \textbf{124,267} \\

        \bottomrule
    \end{tabularx}

    \vspace{1pt}
\end{wraptable}

To collect data from these marketplaces, we develop an automated collection pipeline with four stages: (i) entry collection, (ii) repository validation, (iii) repository deduplication, and (iv) source-code retrieval. Because MCP marketplaces expose heterogeneous interfaces and page structures, we implement source-specific collectors to enumerate server entries through public APIs, SDKs, web pages, or structured data embedded in those pages.
We then normalize project names, descriptions, source-provided categories and tags, repository links, and provenance metadata into a unified JSONL schema, producing the \emph{raw collection}.
Next, we validate associated repositories using the GitHub REST API, excluding inaccessible or empty repositories and documentation-only projects without executable source code. For the remaining entries, we resolve repository URLs to canonical repository identifiers and deduplicate records that refer to the same repository. Finally, we retrieve the source code of each unique repository for subsequent source-level analysis.

\noindent\textbf{Results.}
Table~\ref{tab:mcp_collection} summarizes the collection results. We collect 414,733 raw marketplace records, of which 368,754 are associated with repositories that pass our validity checks (e.g., schema-compliant check). After repository-level deduplication and source-code retrieval, the dataset contains 124,267 unique projects with accessible codebases. Each project retains its source provenance, collection timestamp, original repository link, and normalized repository address to support traceability and reproducibility.

\begin{figure*}[t]
  \centering
  \includegraphics[width=\linewidth]{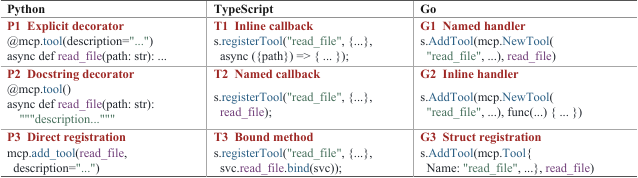}
  \caption{MCP tool registration patterns in Python, TypeScript, and Go.}
  \label{fig:mcp-tool-patterns-read-file}
  \vspace{-15pt}
\end{figure*}

\subsection{Tool Specification Extraction}
\label{sec:extraction}
To support functional taxonomy construction, we extract tool descriptions and function signatures from MCP server source code. This is challenging because MCP tools are implemented differently across programming languages, SDKs, and frameworks. As illustrated in Figure~\ref{fig:mcp-tool-patterns-read-file}, tools may be defined through decorators or annotations, explicit registration calls, tool objects or builders, wrappers or factories, or low-level protocol handlers. Moreover, tool metadata and execution logic may reside in different parts of the codebase.
We therefore develop a cross-language static analysis pipeline that recognizes diverse tool declaration patterns and resolves them to their corresponding implementation functions.

\noindent\textit{(i) Language scope and pattern identification.}
We first analyze the distribution of primary implementation languages across all 124,267 source-available server projects in our collection. We then select the seven most prevalent languages for tool extraction: Python, TypeScript, JavaScript, Go, Rust, Java, and C\#. In total, these languages cover 110,482 repositories (88.9\%). For each selected language, we manually inspect tool implementations and iteratively identify tool declaration and registration patterns. This analysis yields 56 patterns spanning decorators and annotations, explicit registration calls, tool objects and builders, wrappers and factories, and low-level protocol handlers. 
\setlength{\intextsep}{2pt}

\begin{wraptable}{r}{0.46\textwidth}
    \centering

    \setlength{\abovecaptionskip}{0pt}
    \setlength{\belowcaptionskip}{3pt}

    \caption{Scale of MCP tool extraction across seven programming languages.}
    \label{tab:mcp_tool_extraction}

    \scriptsize
    \setlength{\tabcolsep}{14pt}
    \renewcommand{\arraystretch}{1.03}

    \begin{tabular}{@{}lrr@{}}
        \toprule
        \textbf{Language}
        & \textbf{\#Servers}
        & \textbf{\#Tools} \\
        \midrule

        Python
        & 47,939 (43.4\%)
        & 553,615 (41.7\%) \\

        TypeScript
        & 36,103 (32.7\%)
        & 561,612 (42.3\%) \\

        JavaScript
        & 15,808 (14.3\%)
        & 95,308 (7.2\%) \\

        Go
        & 4,183 (3.8\%)
        & 47,435 (3.6\%) \\

        Rust
        & 2,833 (2.6\%)
        & 40,512 (3.1\%) \\

        Java
        & 1,860 (1.7\%)
        & 9,556 (0.7\%) \\

        C\#
        & 1,756 (1.6\%)
        & 20,195 (1.5\%) \\

        \midrule

        \textbf{Total}
        & \textbf{110,482}
        & \textbf{1,328,233} \\

        \bottomrule
    \end{tabular}

    \vspace{-6pt}
\end{wraptable}

\noindent\textit{(ii) Tool recognition and entry-function localization.}
We develop language-specific static analyzers that recognize these patterns in abstract syntax trees (ASTs)~\cite{falleri2024fine,sun2023abstract,dyer2014mining}. For each recognized tool declaration or registration, the analyzers recover its metadata and resolve the associated handler, callback, or method reference to an \emph{entry function} that implements the tool's execution logic. For low-level protocol implementations, the analyzers link tool declarations to invocation-dispatch logic and follow branches selected by tool name to locate the corresponding entry functions. We also analyze the registration patterns, AST backends, and entry-function resolution rules.
We represent each extracted tool using five normalized fields: \texttt{server\_name}, \texttt{tool\_name}, \texttt{description}, \texttt{signature}, and \texttt{return}. This unified representation provides a consistent schema for subsequent functional classification and comparison.

\noindent\textbf{Results.}
As shown in Table~\ref{tab:mcp_tool_extraction}, we extract 1{,}328{,}233 tool specifications across seven languages. Python and TypeScript account for 76.1\% of MCP servers in these seven languages and contribute the majority of extracted tools (84.0\%).

\subsection{Functional Taxonomy Construction}
\label{sec:taxonomy}

\subsubsection{Problem Formulation and Design Goals}
\label{sec:taxonomy-problem}
Given the eligible tool set $\mathcal{T}$, we aim to construct a rooted functional taxonomy $\mathcal{G}$ and assign each tool to exactly one leaf. Each internal node $v$ partitions its tool set $T_v$ along a single \emph{splitting dimension} $\delta_v$, i.e., the functional property that distinguishes its children, such as the target service, the resource, or the operation. Each child category is defined by a name, a description, and explicit inclusion and exclusion criteria. Partitioning continues until each leaf groups \emph{functionally comparable} tools: tools that accomplish the same primary user-level task under comparable inputs and conditions and can therefore substitute for one another. For example, \texttt{execute\_sql}, \texttt{run\_sql}, and \texttt{execute\_query} belong to the same leaf if each executes a user-provided SQL query against a database, despite differences in naming or implementation.

The core decision at each node is which dimension to split on and how to define categories so that every tool fits exactly one.
Neither embedding clustering nor an LLM alone makes this decision well. Clustering groups similar tools but cannot name the
dimension that separates the groups; an LLM can propose meaningful dimensions but may not fit the tools that actually exist. We therefore let an LLM propose each node's design from sampled tools and test it on other sampled tools, guided by three design goals. \textbf{(R1)} Categories describe what a \emph{tool} does, independent of the server or marketplace that lists it.
\textbf{(R2)} Every category design is tested on real tools before it is used. \textbf{(R3)} The procedure scales to over a million tools, so per-tool LLM calls are the exception rather than the rule.

\subsubsection{Construction Methodology}
\label{sec:taxonomy-overview}
The functional taxonomy construction proceeds in five stages.
(\ding{202}) \textit{Taxonomy initialization}: prepares the tool records and seeds the first level of the tree with 18 domains;
(\ding{203}) \textit{Recursive node expansion}: grows the tree top down on a development set by designing, testing, and finalizing one node at a time;
(\ding{204}) \textit{Hybrid routing}: assigns tools to the categories of each finalized node using embedding and LLM-based routing;
(\ding{205}) \textit{Full corpus mapping}: maps the remaining corpus onto the constructed tree; and
(\ding{206}) \textit{Human refinement}: resolves tools that remain unassigned or ambiguous after automated mapping.

\noindent\textbf{\ding{202} Taxonomy initialization.}
We remove tools with a missing or unreadable description or missing signature, leaving
1{,}056{,}806 eligible tools. For routing, each tool $t$ is embedded as
$\mathbf{e}_t = E_\theta(\ d_t \Vert
\texttt{Function: } f_t) \in \mathbb{R}^d$, where $d_t$ is its description,
$f_t$ its signature, and $E_\theta$ is OpenAI's \texttt{text-embedding-3-small}.
Users already browse MCP servers by marketplace category, so we
consolidate the category systems of the 17 marketplaces into 18 domains,
which form the first level of the tree. Because marketplace labels describe
servers rather than tools (R1), we use them only as seed vocabulary and as
sampling strata, and assign each tool to a domain individually. The root is
processed by Algorithm~\ref{alg:expand-node} with one difference: the
consolidated design replaces the Designer's proposal (line~\ref{ln:design})). 
Designing every node from all 1.06M tools would require LLM calls over the
whole corpus at every level (R3). We therefore build the tree on a
\emph{development set} $\mathcal{T}_{\text{dev}}$: 5{,}000 servers, sampled
in proportion to their domains, together with all 75{,}365 of their tools,
which we call \emph{development tools}. All other tools are mapped onto the
finished tree in \ding{205}. 

\begin{wrapfigure}{r}{0.5\textwidth}
\vspace{0pt}
\begin{minipage}{\linewidth}
\scriptsize
\algrenewcommand\alglinenumber[1]{\scriptsize #1}
\hrule height 0.8pt\kern 2pt
\captionsetup{type=algorithm, font=scriptsize, labelfont={bf,scriptsize},
  textfont=scriptsize, justification=raggedright, singlelinecheck=false,
  skip=0pt, position=top}
\captionof{algorithm}{Expanding node $v$. $\hat p(S)$ and $F(S)$ are the
failure rate and the failed tools of the pilot sample $P$ under design $S$;
\textsc{Accept} is the pilot acceptance test.}
\label{alg:expand-node}
\kern 2pt\hrule\kern 2pt
\begin{algorithmic}[1]
\Require tools $T_v$, path context $\Pi(v)$
\Ensure frozen design $S_v$ and child tool sets, or leaf
\State \textbf{if} $|T_v| \le 1$ \textbf{then return} leaf\label{ln:leaf}
\State \textbf{if} $|T_v| < n_{\min}$ \textbf{and} \label{ln:2}
  \Call{Comparable}{$T_v$} \textbf{then return} leaf
\State $(D,P) \gets \Call{DisjointSample}{T_v,n_d,n_p}$\label{ln:sample}
\State $S \gets \Call{Designer}{D,\Pi(v)}$\label{ln:design}
\State \textbf{if} $S$ is terminal \textbf{and}
  \Call{Comparable}{$T_v$} \textbf{then return} leaf
\State \textbf{if} $S$ is terminal \textbf{then}
  $S \gets \Call{Designer}{D,\Pi(v),\text{leaf feedback}}$
\For{$r \gets 0$ \textbf{to} $K$}\label{ln:loop}
  \State $A \gets \Call{Classifier}{S,P}$
  \State $F(S) \gets
    \{t \in P : A_t=\varnothing \lor \gamma_t<\tau_c\}$
  \State $\hat p(S) \gets |F(S)|/|A|$
  \State \textbf{if} \Call{Accept}{$\hat p(S),|T_v|$}
    \textbf{then break}\label{ln:gate}
  \State \textbf{if} $r<K$ \textbf{then}
    $S \gets \Call{Refiner}{S,F(S)}$\label{ln:refine}
\EndFor\label{ln:13}
\State $\tilde S \gets \Call{MECEValidator}{S,P,A}$\label{ln:validate}
\State \textbf{if} $\tilde S$ is schema-valid
  \textbf{then} $S \gets \tilde S$\label{ln:repair}
\State $S_v \gets S$; flag $v$ \textbf{if not}
  \Call{Accept}{$\hat p(S_v),|T_v|$}\label{ln:freeze}
\State route $T_v$ by Eq.~\ref{eq:vector-routing},
  with Classifier fallback\label{ln:route}
\State failed or low-confidence tools $\rightarrow U$\label{ln:residual}
\State \Return non-empty children of $v$\label{ln:return}
\end{algorithmic}
\kern 1pt\hrule height 0.8pt
\end{minipage}
\vspace{2pt}
\end{wrapfigure}

\noindent\textbf{\ding{203} Recursive node expansion.}
Starting from the 18 domains, nodes are expanded layer by layer. Independent nodes are processed concurrently, and each materialized child becomes immediately eligible for expansion without waiting for the rest nodes.
Algorithm~\ref{alg:expand-node} lists steps for one node $v$.

\underline{\textit{Leaf test} (line~\ref{ln:leaf}--\ref{ln:2})}.
A singleton node is immediately accepted as a leaf. If \(v\) contains fewer
than \(n_{\min}\) tools, an LLM-based Leaf Validator examines a sample of up
to \(n_{\ell}\) tools, and \(v\) becomes a leaf only when those tools form a
functionally comparable set. Otherwise, the node continues to category
design despite being smaller than \(n_{\min}\). A terminal proposal returned
by the Designer is subject to the same validation; if it is rejected, the
Designer is called again with the validator's feedback. Thus,
\(n_{\min}\) triggers leaf validation rather than unconditionally terminating
the branch.

\underline{\textit{Sampling} (lines~\ref{ln:sample})}.
Category design and pilot validation use disjoint samples. Tools are placed
in a deterministic hash-based order and divided into a design sample \(D\)
of at most \(n_d\) tools and a pilot sample \(P\) of at most \(n_p\) tools.
When the node contains fewer than \(n_d+n_p\) tools, the available tools are
divided proportionally while preserving at least one pilot tool whenever
possible. Consequently, no pilot tool is exposed to the Designer before
validation.

\underline{\textit{Category design} (line~\ref{ln:design})}.
A Designer LLM receives the design sample and the \emph{path context}
$\Pi(v)$, i.e., the dimension and category of every ancestor of $v$. It
proposes one splitting dimension that is finer than, and different from,
all ancestor dimensions, together with the child categories. Deterministic
checks reject malformed designs, duplicate categories, and repeated
dimensions; they apply to every design, including those later returned by
the Refiner and the Validator.

\underline{\textit{Pilot validation and refinement} (lines~\ref{ln:loop}--\ref{ln:13})}.
A Classifier LLM assigns each pilot tool to one category, or to none if no
definition fits, and reports a confidence $\gamma \in [0,1]$. A tool
\emph{fails} if it fits no category, which indicates a coverage gap, or if
$\gamma < \tau_c$, which indicates an unclear boundary; only malformed
outputs are retried. 
Let $\hat p$ be the fraction of failed pilot tools. A design passes the
acceptance test, \textsc{Accept}$(\hat p)$, if \[
\hat p(S)\leq\varepsilon
\quad\text{and}\quad
\hat p(S)\,|T_v|\leq\eta .
\], where $|T_v|$ is the number of tools currently assigned to node \(v\). The first condition bounds the failure
\emph{rate}; the second bounds the expected \emph{number} of failures,
which matters for large nodes. 
If the design fails the test, a Refiner LLM receives the failed tools, the
Classifier's explanations, and current splitting design. It
applies the least disruptive fix first: (1)~add categories, (2)~revise
definitions, or (3)~replace the splitting dimension, and moves to the next
fix when the previous one did not lower $\hat p$. Every revision is tested
on the \emph{whole} pilot sample, because a changed definition can also
move tools that passed before. After at most \(K\) refinement rounds, we retain the latest schema-valid design. 

\underline{\textit{Node finalization} (lines~\ref{ln:validate}--\ref{ln:return})}.
A Validator LLM checks the best design against the pilot tools assigned to
each category: categories must not overlap, must jointly cover all tools,
must follow one dimension, and must have clear boundaries and comparable
granularity. The Validator may return a repaired design, which we accept
only if it passes the same deterministic schema, category-naming, and
dimension-progression checks applied to Designer and Refiner outputs
(line~\ref{ln:repair}). We then freeze the design and flag $v$ if it never
passed the acceptance test. Pilot tools keep their Classifier labels, all
other tools at $v$ are assigned to children by hybrid routing (\ding{204}),
and tools that remain unplaced, including pilot tools that fail under the
frozen design, join the residual set $U$ (\ding{206}). The non-empty
children are expanded in turn. 

\noindent\textbf{\ding{204} Hybrid routing.}
Hybrid routing assigns the tools of a finalized node to its children, both
during construction (line~\ref{ln:route}) and during corpus mapping. An LLM
call per tool is too expensive at our scale (R3), so we use embedding
similarity whenever it is decisive. Let $k_1$ and $k_2$ be the two
categories whose embeddings $\mathbf{c}_k$ (computed from name, definition,
and inclusion criteria) are most similar to $\mathbf{e}_t$. We accept $k_1$
if
\begin{equation}
  \operatorname{sim}(\mathbf{e}_t,\mathbf{c}_{k_1}) \ge \tau_s(\ell)
  \;\;\text{and}\;\;
  \operatorname{sim}(\mathbf{e}_t,\mathbf{c}_{k_1})
  - \operatorname{sim}(\mathbf{e}_t,\mathbf{c}_{k_2}) \ge \tau_\Delta(\ell),
  \label{eq:vector-routing}
\end{equation}
i.e., if $k_1$ is both close to the tool and clearly ahead of the
runner-up; otherwise, the Classifier decides. Embeddings separate topics
better than operations (e.g., \texttt{create\_issue} and
\texttt{delete\_issue} are close), so the thresholds depend on the tree
level $\ell$: for each level, we choose the most permissive thresholds
whose vector assignments agree with the Classifier's pilot labels on at
least a fraction $\pi$ of tools. A tool that the Classifier cannot place,
or places with $\gamma < \tau_c$, stays at $v$ and joins $U$.

\noindent\textbf{\ding{205} Full corpus mapping.}
Development tools keep their construction assignments. Every other
eligible tool is first matched against embeddings of complete root-to-leaf
paths (the names and definitions along each path), using
Eq.~\ref{eq:vector-routing} with separately calibrated thresholds. If this
match is not decisive, the tool is routed top-down by \ding{204}, with the
Classifier deciding at the node where Eq.~\ref{eq:vector-routing} fails.
A leaf that looked homogeneous on a few development tools may receive many
diverse corpus tools, after full-corpus mapping, unresolved and low-confidence tools are collected
for post-classification refinement and human review. 
Finally, because each node chooses its dimension locally, different
branches reach the same level of specificity at different depths. An LLM
therefore reviews a sample of root-to-leaf dimension sequences and derives
an ordered set of \emph{semantic ranks}; each dimension receives the rank
it holds in most paths that contain it. This level alignment only labels
nodes and leaves all memberships unchanged; cross-domain comparisons in
RQ1 use ranks instead of raw depth. 

\noindent\textbf{\ding{206} Human refinement.}
The residual set $U$ contains the tools that the Classifier could not
place, or placed with low confidence, during construction or corpus
mapping. Even a small residual fraction of a million tools is too large to
label exhaustively, so we resolve $U$ in rounds. In each round, reviewers
label 200 tools sampled from $U$ across the nodes at which they failed:
each tool is assigned to an existing category, marked as a capability
missing under that node, or marked as not a functional tool. We use these
labels in two ways: (i)~we add rules and labeled examples to the
Classifier's instructions for the affected nodes, and (ii)~when several
tools share a missing capability, we add a new leaf category, which must
pass the same checks as any design (category design and node
finalization). Existing nodes are never deleted, merged, or moved, so
earlier assignments remain valid. The Classifier then re-classifies the
tools remaining in $U$. Refinement stops when $U$ holds fewer than 0.5\% of
all tools or after $R_{\max}$ rounds. Review labels are excluded from all
evaluation data. 

\subsubsection{Implementation}
\label{sec:taxonomy-impl}
The Designer, Classifier, Refiner and the Validator all uses gpt-5.6-luna. We set
$\tau_c = 0.55$, $\varepsilon = 0.05$, $\eta = 50$, $K = 3$, $n_{\min}=100$, and $R_{\max=10}$. $n_p$ and $n_d$ are
both $(1000,1000,500,200,100,50)$, with the final value used for all
subsequent depths. Up to 64 workers run design,
refinement, and validation, and 50 workers run classification.

\subsubsection{Taxonomy Quality Evaluation}
\label{sec:taxonomy-eval}
Our analyses rely on two properties of the taxonomy: tools are placed in the correct leaf, and tools that share a leaf are functionally comparable.
We evaluate both on a uniform random sample of 200 tools with a leaf assignment. For each sampled tool, we ask (i)~whether it belongs to its assigned leaf (\emph{assignment precision}), and (ii)~whether it is functionally comparable to a randomly chosen tool from the same leaf (\emph{leaf comparability});  We use an LLM judge (gpt-5.6-terra), a different model than the models used for construction, to answer both questions given the tools' specifications, the root-to-leaf path with its category definitions, and our definition of functional comparability. To validate the judge, two authors independently labeled a random subset of 100 tools without seeing the judge's answers and resolved disagreements by discussion.

\noindent\textbf{Results.}
90.19\% of the eligible tools reach a leaf. Of the sampled tools, 87.00\% belong to their assigned leaf, and 85.00\% of same-leaf pairs are functionally comparable. 
The judge matches two authors' consolidated labels on 95.40\% of assignment and 94.12\% of comparability questions ($\kappa = $ 0.82 and 0.85). Most errors involve target or operation boundaries (e.g., tools placed in a sibling leaf, acting on the same resource but performing a different operation). These results indicate that leaf membership is a reliable basis for the analysis in Section ~\ref{sec:measurement}.

\section{Measurement}
\label{sec:measurement}

\subsection{RQ1: Functional landscape.}
Prior MCP studies characterize the ecosystem at the server level, mostly
through marketplace-defined categories~\citep{guo2025measurement,lin2025mcpcorpus}.
Server-level labels, however, cannot reveal which operations agents can
actually invoke, how many distinct tool capabilities exist, or how finely these
capabilities must be distinguished. Using the taxonomy as a measurement
instrument, RQ1 examines which domains MCP tools serve (Finding~1), whether
large domains also offer more diverse capabilities (Finding~2), and how deep the taxonomy must go before tools in a domain become functionally comparable. The analysis covers the 953{,}172 tools assigned to one of 58{,}915 leaf capabilities.

\begin{figure}
    \centering
    \includegraphics[width=1.0\linewidth]{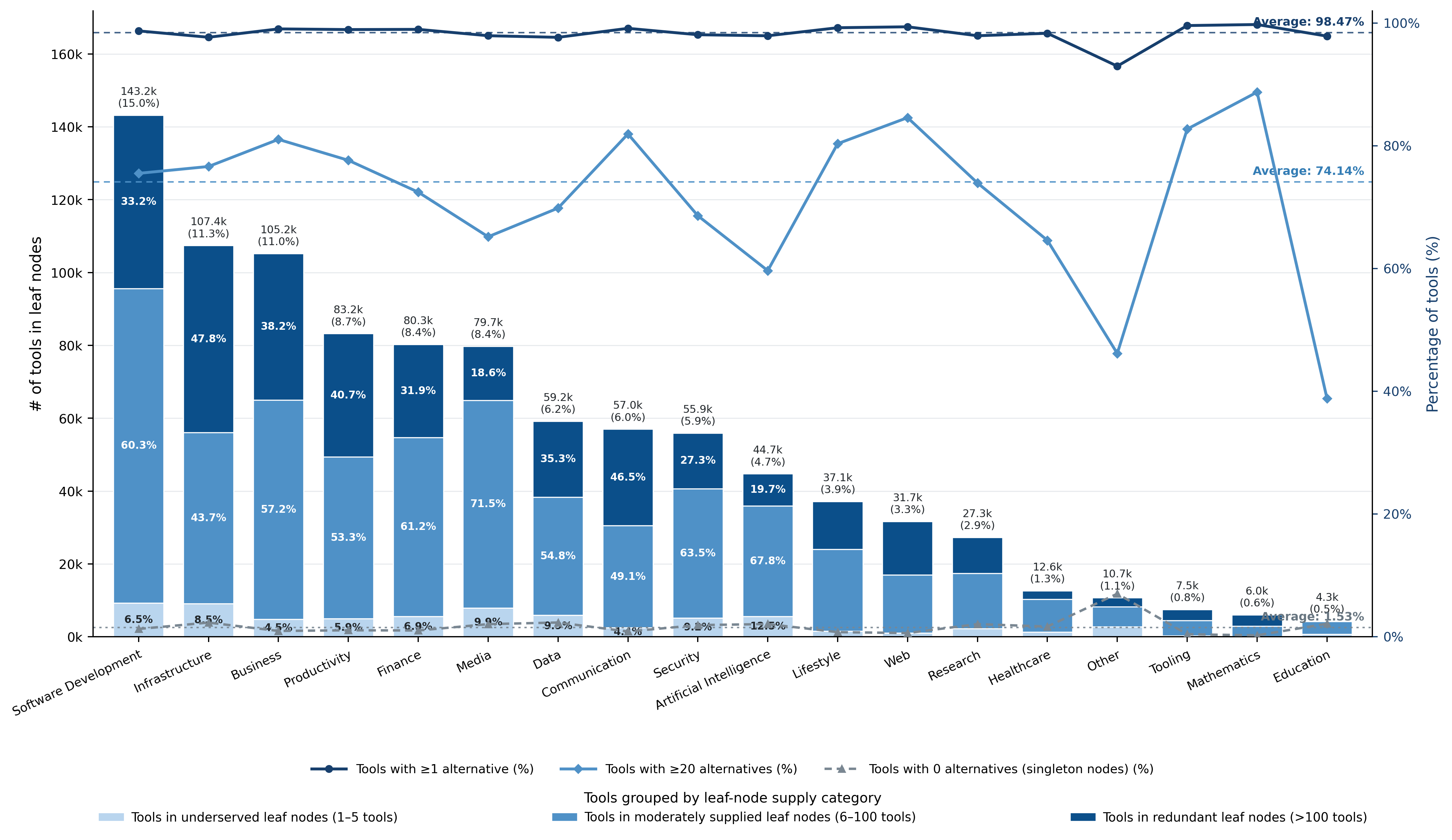}
    \caption{Functional distribution, fragmentation, and candidate-alternative availability across the MCP ecosystem. Bar height denotes the total number of tools in leaf nodes. Stacked segments divide tools according to the number of candidate functional alternatives in the same leaf node. The lines show the percentages of tools with at least one, at least 20, and no candidate alternatives, respectively.}
    \label{fig:functional_landscape}
\end{figure}

\vspace{2pt}\noindent\textbf{Finding 1: MCP has outgrown software development: 85\% of tools serve other domains.}
Figure~\ref{fig:functional_landscape} shows that Software Development is the largest individual domain with 143{,}202 tools (15.0\%), followed by Infrastructure (11.3\%) and Business (11.0\%). The remaining of tools span domains including Finance, Media, Communication, Healthcare, and Education. These results show that MCP is no longer primarily a developer-tool ecosystem; it has become a general-purpose integration layer covering a wide range of application domains. Consequently, prior benchmarks built from a few dozen curated servers~\citep{wang2026mcp,bandi2026mcp,luo2025mcp} therefore cover only a small part of the functional range that deployed agents face.

\begin{wraptable}{r}{0.50\textwidth}
\centering
\caption{Within-domain concentration in representative domains.}
\label{tab:domain-concentration}
\renewcommand{\arraystretch}{0.95}
\resizebox{\linewidth}{!}{%
\begin{tabular}{lrrrr}
\toprule
\textbf{Domain} &
\multicolumn{2}{c}{\textbf{Largest leaf}} &
\multicolumn{2}{c}{\textbf{Top five leaves}} \\
\cmidrule(lr){2-3}\cmidrule(lr){4-5}
&
\textbf{Tools} &
\textbf{Share (\%)} &
\textbf{Tools} &
\textbf{Share (\%)} \\
\midrule
Software Development  & 1,128 & 0.8  & 4,794 & 3.3  \\
Business              &   826 & 0.8  & 3,529 & 3.4  \\
Artificial Intelligence & 481 & 1.1  & 1,870 & 4.2  \\
Infrastructure        & 2,790 & 2.6  & 9,373 & 8.7  \\
Education             &   124 & 2.9  &   431 & 10.0 \\
Tooling               &   569 & 7.6  & 1,700 & 22.7 \\
Mathematics           &   934 & 15.5 & 2,085 & 34.6 \\
\bottomrule
\end{tabular}%
}

\end{wraptable}
\vspace{2pt}\noindent\textbf{Finding 2: More tools do not mean more capabilities: large domains fragment, while smaller domains often repeat the same capabilities.}
A domain can be large for two different reasons: it offers many distinct \emph{capabilities} (leaf categories of functionally comparable tools), or many tools implement the same few capabilities. Because these cases pose different problems for users when retrieving tools, we ask whether tool volume reflects capability breadth or redundancy.

For each domain, we count its capabilities and measure the share of its tools in its largest and five largest capabilities (Table \ref{tab:domain-concentration}).
Software Development spans 8{,}213 capabilities, yet its largest one holds only 0.8\% of its tools and its five largest 3.3\%; Business is similarly dispersed (0.8\% and 3.4\%).
In contrast, Mathematics has only 154 capabilities, and its five largest hold 34.6\% of its tools; Tooling shows the same pattern (295 capabilities; 22.7\%).
Tool volume thus reflects breadth in some domains and redundancy in others, which creates two distinct challenges.
In fragmented domains, the challenge is \emph{discovery}: users must locate one of thousands of narrow capabilities, which coarse server-level categories cannot support.
In concentrated domains, the challenge is \emph{selection} among many tools that implement the same capability, which we examine in RQ2 and RQ4.

\vspace{2pt}\noindent\textbf{Finding 3: Functional granularity is domain-dependent: most capabilities are resolved within five to six distinctions, but AI, Web, and Data carry a long tail that requires eight or more.}
Finding~2 shows how \emph{broad} each domain is; we next ask how \emph{fine} its distinctions are, i.e., how deep the taxonomy must go before tools become functionally comparable.
This matters because marketplaces organize servers into a fixed, shallow list of categories: if domains need different numbers of distinctions, any fixed-depth scheme is too coarse for some domains and unnecessarily fine for others.
\begin{wrapfigure}{r}{0.5\textwidth}
    \centering
    \includegraphics[width=0.9\linewidth]{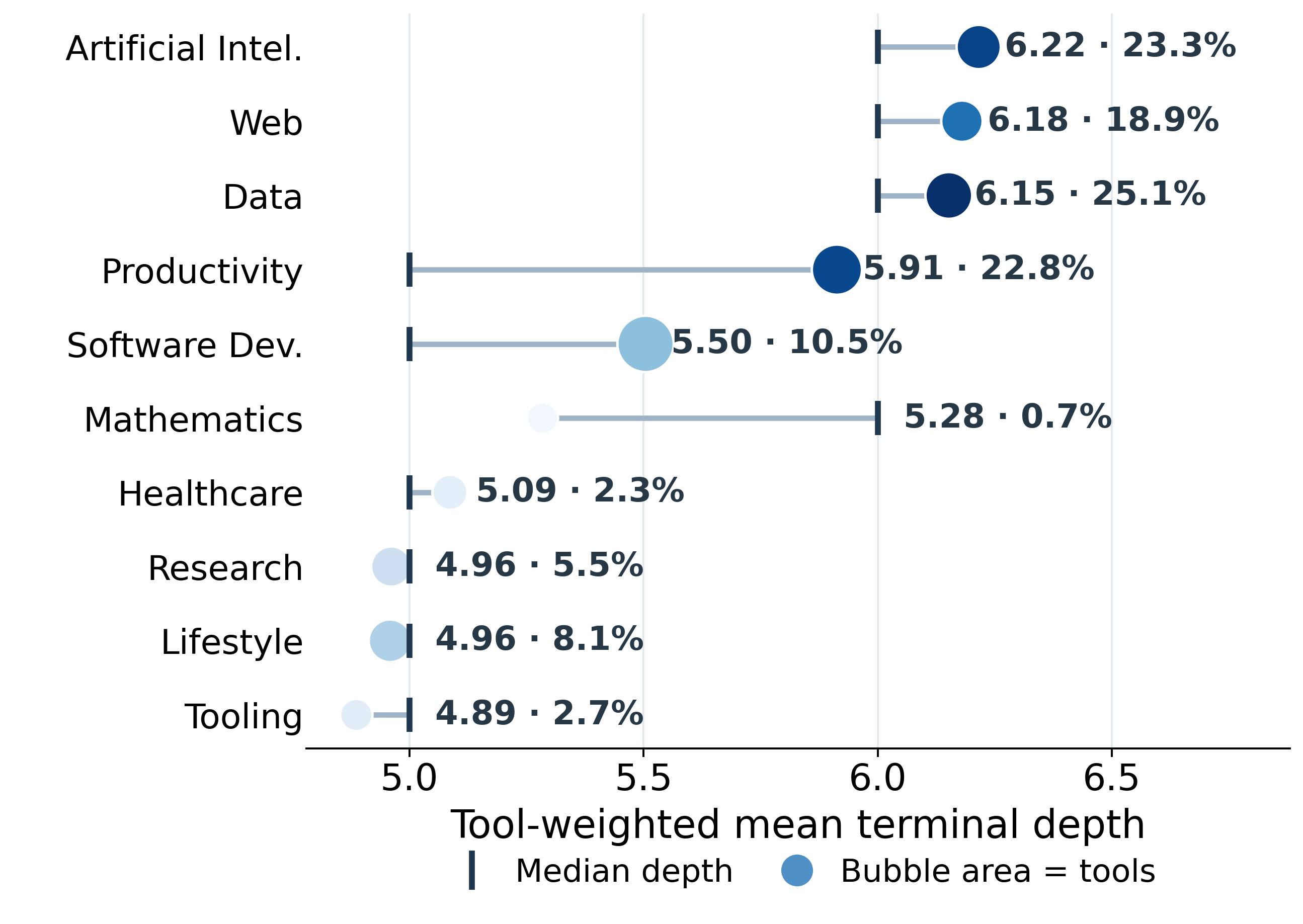}
    \caption{Specialization depth per domain. Bubble position is the
    tool-weighted mean terminal depth, vertical markers the median; bubble
    size is the number of tools, and color the share of tools at depth 8 or
    deeper.}
    \label{fig:domain_depth}
\end{wrapfigure}
For each domain, we measure leaf depth, i.e., the number of splits from the root, and the share of tools at depth 8 or deeper (Figure~\ref{fig:domain_depth}). Median depth is consistently 5 or 6 across domains.
The domains differ in their tails. In Artificial Intelligence, Web, and Data (mean depth 6.15--6.22), 18.9--25.1\% of tools sit at depth 8 or deeper, as do 22.8\% in Productivity; in Tooling, Lifestyle, and Research (mean 4.89--4.96), this share is at most 8.1\%.
Mathematics shows a third profile: a median depth of 6 but only 0.7\% of tools at depth 8 or deeper, i.e., uniformly specialized capabilities without a long tail.
These long tails suggest that the deep domains combine many providers, resources, and operation variants, so some of their capabilities are identified only after many successive distinctions.
Tool organization and retrieval should therefore adapt their granularity per domain, and even per branch, rather than impose a uniform depth.

\answerbox{RQ1}{The MCP ecosystem is functionally broad with 18 domains and 58{,}915 distinct capabilities. Yet this breadth is uneven: a few general-purpose domains dominate tool volume; Large domains span many narrow capabilities, whereas some smaller domains concentrate on a few. Users need a tool-level map with domain-adaptive depth to discover capabilities and choose among functionally comparable tools.}

\subsection{RQ2: Functional Comparability and Supply}
\label{sec:rq2}
When a tool is unavailable, deprecated, or insecure, an agent can only
recover if another tool performs the same operation. Conversely, if many
developers implement the same capabilities while others remain uncovered,
effort is duplicated, and parts of the ecosystem are fragile. RQ2 uses leaf
co-membership as a proxy for functional comparability and examines how many alternative tools have (Finding~4) and how tool supply
is distributed across capabilities (Finding~5).

\vspace{2pt}\noindent\textbf{Finding 4: Alternatives are the norm: 98.5\% of tools have at least one functionally comparable alternative, and 74.1\% have at least 20.}
To assess whether agents and users can routinely choose among tools with comparable capabilities, we count, for each tool, its candidate alternatives, i.e., the other tools in the same taxonomy leaf, and report the share of tools with none, at least one, and at least 20 alternatives, overall and per domain (lines in Figure~\ref{fig:functional_landscape}).
Only 1.53\% of tools have no alternative, whereas 98.47\% have at least one and 74.14\% have at least 20.
The share with at least one alternative is high in every domain; domains differ mainly in \emph{how many} alternatives tools have: Mathematics, Tooling, and Web have the highest shares of tools with at least 20 alternatives, whereas fewer than half of the tools in Education and Other do.
For most tools, the open question is therefore not whether an alternative exists but which one to choose, which motivates comparing how alternatives differ (RQ3) and organizing retrieval around comparable candidates (RQ4).

\noindent\textbf{Finding 5: Nearly one quarter of observed capabilities
have no alternative.}
Although alternatives are common among tools (Finding~4), 58,915 capabilities (24.80\%) contain exactly one observed tool. We call these \emph{single-tool capabilities}. To understand the support behind their sole implementations, we examine their domain distribution and the characteristics of their supporting repositories.
Single-tool capabilities are most prevalent in Other and Infrastructure, representing 42.16\% and 35.40\% of capabilities within these domains, respectively. Examples include Cryptographic Proofs and Pronunciation Dictionary Management.
Since agents have no fallback for these capabilities, we examine whether their sole implementations are well supported. We compare repositories that support at least one single-tool capability with repositories whose capabilities all have alternatives, counting each repository once, in terms of GitHub stars, recent activity (a push within 90 days before collection), and owner type.
Repositories supporting single-tool capabilities are more often recently
active (47.50\% versus 39.26\%) and organization-owned (18.83\% versus
14.61\%), differences of 8.24 and 4.22 percentage points, respectively.
Their median star count is slightly higher (2 versus 1). 
Thus, lacking alternatives does not imply weak repository support. Single-tool capabilities are, in fact, backed by more recently active repositories on average. One possible explanation is that some tools serve specialized organizational needs, motivating continued development without attracting alternative implementations.

\answerbox{RQ2}{Functionally comparable alternatives are widespread: almost every tool has at least one, and three in four have at least 20. However, this abundance is uneven across capabilities: nearly one quarter have only a single observed tool. Importantly, lacking alternatives does not imply weak repository support, showing that alternative availability and maintenance are distinct dimensions of the MCP ecosystem.}

\subsection{RQ3: Alternative heterogeneity.} 
RQ2 shows that most tools have functionally comparable alternatives, so users and agents routinely face a choice among them.
Functional comparability, however, says nothing about how each alternative is built or maintained.
RQ3 therefore compares alternatives within the same taxonomy leaf and examines whether they differ in security-relevant implementation properties (Finding~6), code complexity and size (Finding~7), and the characteristics of the projects that host them (Finding~8).

\setlength{\intextsep}{2pt}

\begin{wraptable}{r}{0.46\textwidth}
    \centering

    \setlength{\abovecaptionskip}{0pt}
    \setlength{\belowcaptionskip}{3pt}

    \caption{Category distribution of mapped Semgrep alerts.}
    \label{tab:rq3_semgrep_distribution}

    \scriptsize
    \setlength{\tabcolsep}{3.5pt}
    \renewcommand{\arraystretch}{1.03}

    \begin{tabular}{@{}lrrr@{}}
        \toprule

        \textbf{Category}
        & \textbf{\#Alerts}
        & \textbf{\#Tools}
        & \makecell{\textbf{\#Leaf}\\\textbf{Groups}} \\

        \midrule

        Input Validation
        & 1,672 (89.22\%)
        & 939
        & 803 \\

        Transport Security
        & 48 (2.56\%)
        & 42
        & 41 \\

        Resource Management
        & 25 (1.33\%)
        & 24
        & 23 \\

        Credential Exposure
        & 23 (1.23\%)
        & 18
        & 18 \\

        Access Control
        & 3 (0.16\%)
        & 3
        & 3 \\

        Authentication
        & 2 (0.11\%)
        & 2
        & 2 \\

        Other
        & 101 (5.39\%)
        & 62
        & 54 \\

        \bottomrule
    \end{tabular}

    \vspace{2pt}

    \parbox{0.96\linewidth}{
        \tiny
        \textit{Note.} Each alert is mapped to exactly one category;
        tools and leaf-level functional groups may appear in multiple
        categories.
    }

    \vspace{-5pt}
\end{wraptable}
\vspace{2pt}\noindent\textbf{Finding 6: Security risks differ across functionally comparable tools, with improper input validation accounting for most alerts.}

To compare the security-relevant implementations of alternatives, we apply 755 Semgrep rules~\cite{zhang2025automatically} covering the seven target languages to each tool's entry-point function and use an LLM (\texttt{deepseek-flash}) to map the resulting alerts into seven security categories: Input Validation, Transport Security, Resource Management, Credential Exposure, Access Control, Authentication, and Other.
Of the 66{,}912 tools with valid measurements, 1{,}044 (1.56\%) receive at least one of 1{,}874 attributed alerts, spread across 890 functional groups (Table~\ref{tab:rq3_semgrep_distribution}).
Input validation dominates, accounting for 89.22\% of alerts and appearing in 803 groups, whereas no other named category exceeds 2.56\% (the residual Other category accounts for 5.39\%); most alerted tools fall into a single category.

\begin{figure}[htbp]
  \centering
  \begin{minipage}[b]{0.48\linewidth}
    \centering
    \includegraphics[width=1.0\linewidth]{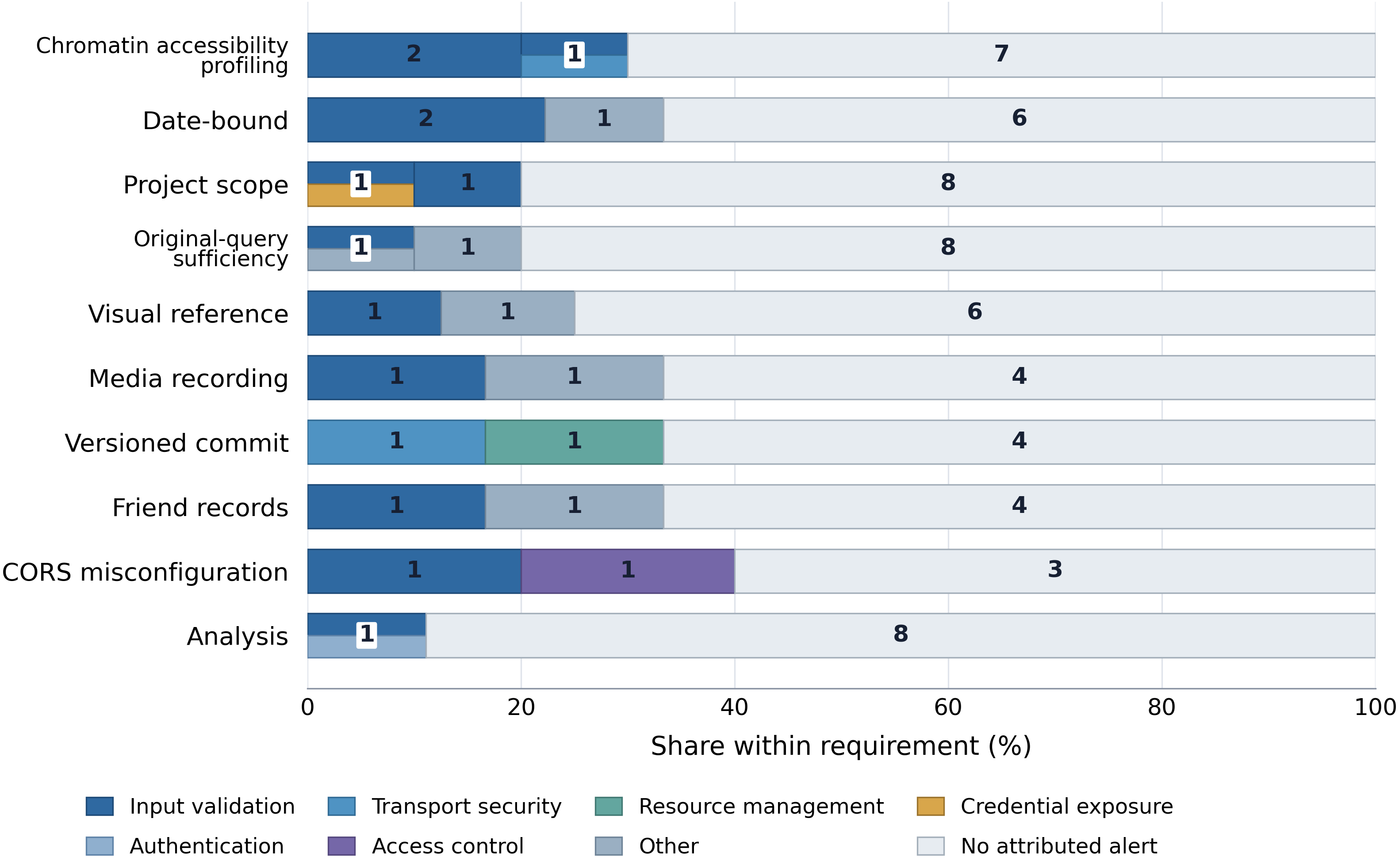}
    \caption{Tool-level Semgrep alerts in ten functional groups. Bars show the share of tools in each group (counts inside bars); colors denote security categories, stacked for tools with several categories, and gray denotes tools without alerts.}
    \label{fig:RQ3_leaf_top10_en}
  \end{minipage}
  \hfill
  \begin{minipage}[b]{0.50\linewidth}
    \centering
    \includegraphics[width=1.0\linewidth]{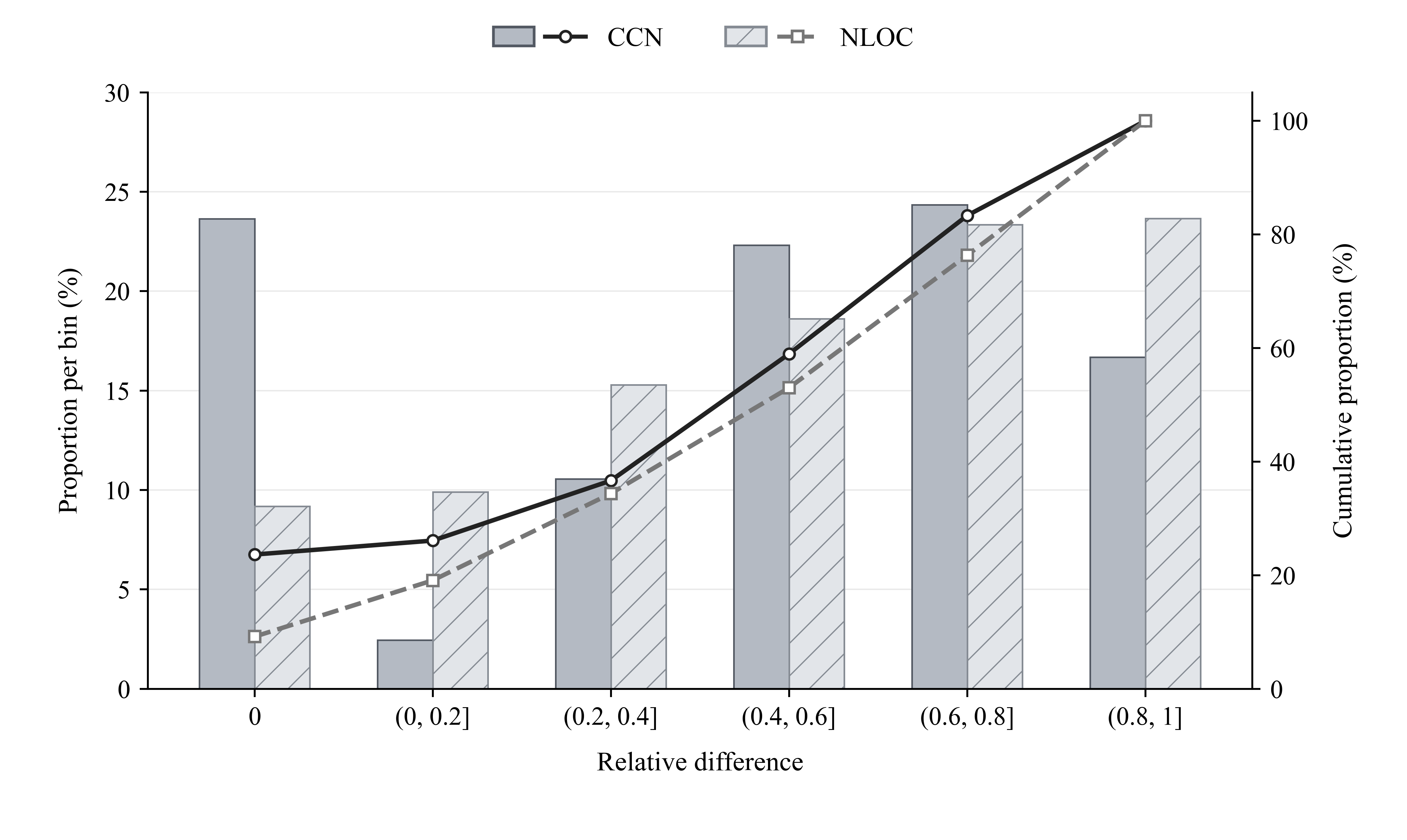}
    \caption{Relative differences in entry-function CCN and NLOC between functionally comparable tools. Bars show the weighted share of tool pairs per difference bin (left axis); lines show the cumulative share (right axis).}
    \label{fig:RQ3_Lizard_pair_combo}
  \end{minipage}
\end{figure}

%
To examine differences within groups, we select ten groups that together cover all security categories (79 tools; Figure \ref{fig:RQ3_leaf_top10_en}).
Every group contains both security alerted and alert-free tools, and security alerted tools make up 11.1--40.0\% of each group; 
Alerted tools in the same group also differ in the kind of risk they carry: in all nine groups with more than one alerted tool, the alerts span at least two security categories.
In \emph{Versioned commit}, for example, one tool sends unencrypted HTTP requests (transport security), while another builds regular expressions dynamically, a potential denial-of-service risk (resource management); in \emph{CORS misconfiguration}, one alerted tool lacks input validation and another lacks access control.
Input validation is the most widespread concern, appearing in nine of the ten groups, and four tools combine it with a second category, such as authentication in \emph{Analysis} or credential exposure in \emph{Project scope}.
Even when alerted tools share a category, the underlying issues can differ: in \emph{Chromatin accessibility profiling}, three tools are flagged for input validation, but two pass unchecked input to subprocess calls, whereas the third builds URLs dynamically.
Functional similarity therefore does not imply similar security posture: in every examined group, users could choose an alert-free tool with the same functionality, and the alerts of the others point to distinct, rule-specific risks.

\begin{figure}
    \centering
    \includegraphics[width=1.0\linewidth, height=6cm]{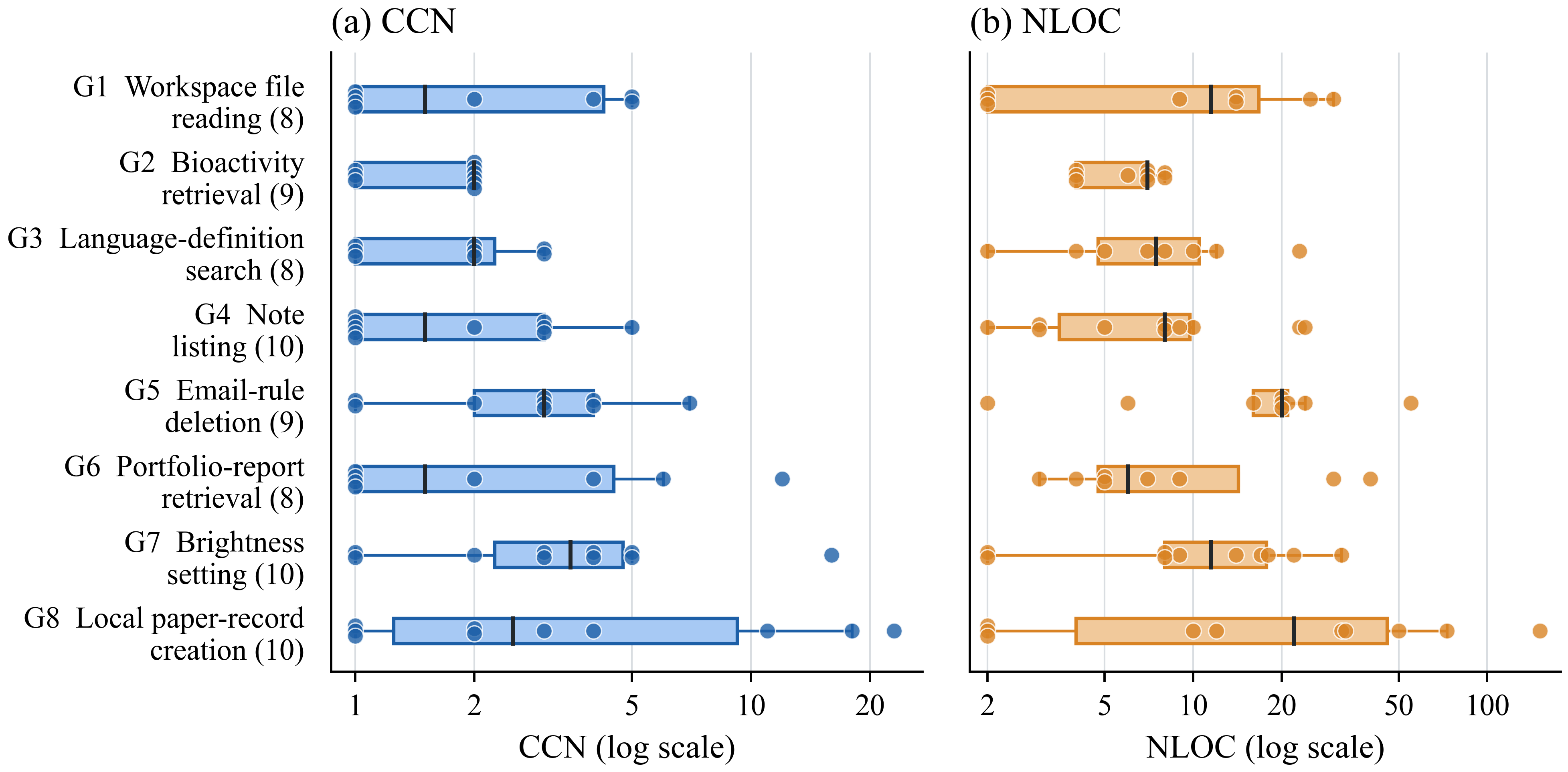}
    \caption{Distributions of entry-function CCN and NLOC across eight functional groups. Numbers in parentheses denote tool counts, and each point represents one tool. Boxes span the interquartile range (IQR), center lines mark the median, and whiskers extend to the most extreme observations within $1.5\times\mathrm{IQR}$ of the box edges. Horizontal axes use logarithmic scales with panel-specific ranges; tick labels show the original measurement values, and all boxplot statistics are computed from the untransformed values.}
    \label{fig:RQ3_Lizard_CCN_NLOC_boxes}
\end{figure}

\vspace{2pt}\noindent\textbf{Finding 7: Comparable tools differ widely in implementation complexity and size, even in the same language: for 41.0\% of pairs.}
To compare the implementation characteristics of alternatives, we use Lizard~\citep{yin2019lizard} to measure the cyclomatic complexity (CCN) and non-comment lines of code (NLOC) of each tool's entry function, two established software metrics for characterizing code complexity and quality~\citep{landman2016empirical,mccabe1976complexity}.\\
To isolate implementation differences from functionality and language, we compare only cross-repository tool pairs within the same leaf and programming language, and define the relative difference of metric $x$ as $d_x(i,j) = |x_i - x_j| / \max(x_i, x_j)$, with $d_x = 0$ if both values are zero; we weight each comparison unit equally so that large groups do not dominate.
The analysis covers 86{,}387 pairs among 55{,}651 tools in 16{,}484 leaf--language units (Figure~\ref{fig:RQ3_Lizard_pair_combo}).
Identical implementations are uncommon: only 23.64\% of pairs have the same CCN and 9.19\% the same NLOC.
Large gaps are common: 41.03\% of pairs differ by $d > 0.6$ in CCN and 47.00\% in NLOC, i.e., the smaller value is less than 40\% of the larger one; for NLOC, 23.66\% of pairs fall in the most extreme bin $(0.8, 1]$.
Within groups, this variation takes two forms (Figure~\ref{fig:RQ3_Lizard_CCN_NLOC_boxes}(a)): in \emph{Brightness Setting}, eight of nine tools have a CCN of at most 5 and a single outlier reaches 16, whereas in \emph{Local Paper Record Creation} the variation spans the whole group (CCN interquartile range 1.25--9.25), and its longest entry function is 75.5 times as long as its shortest (Figure~\ref{fig:RQ3_Lizard_CCN_NLOC_boxes}(b)).

The two metrics are complementary: in \emph{Workspace File Reading}, two tools with the same CCN of 1 contain 2 and 14 lines of code.
Functionally comparable tools thus range from thin wrappers to substantial implementations, so complexity and size offer useful, complementary signals for choosing among alternatives, for example, to estimate review or maintenance effort.\\



%
\begin{wrapfigure}{r}{0.5\textwidth}
    \centering
    \includegraphics[width=1.0\linewidth, height=6cm]{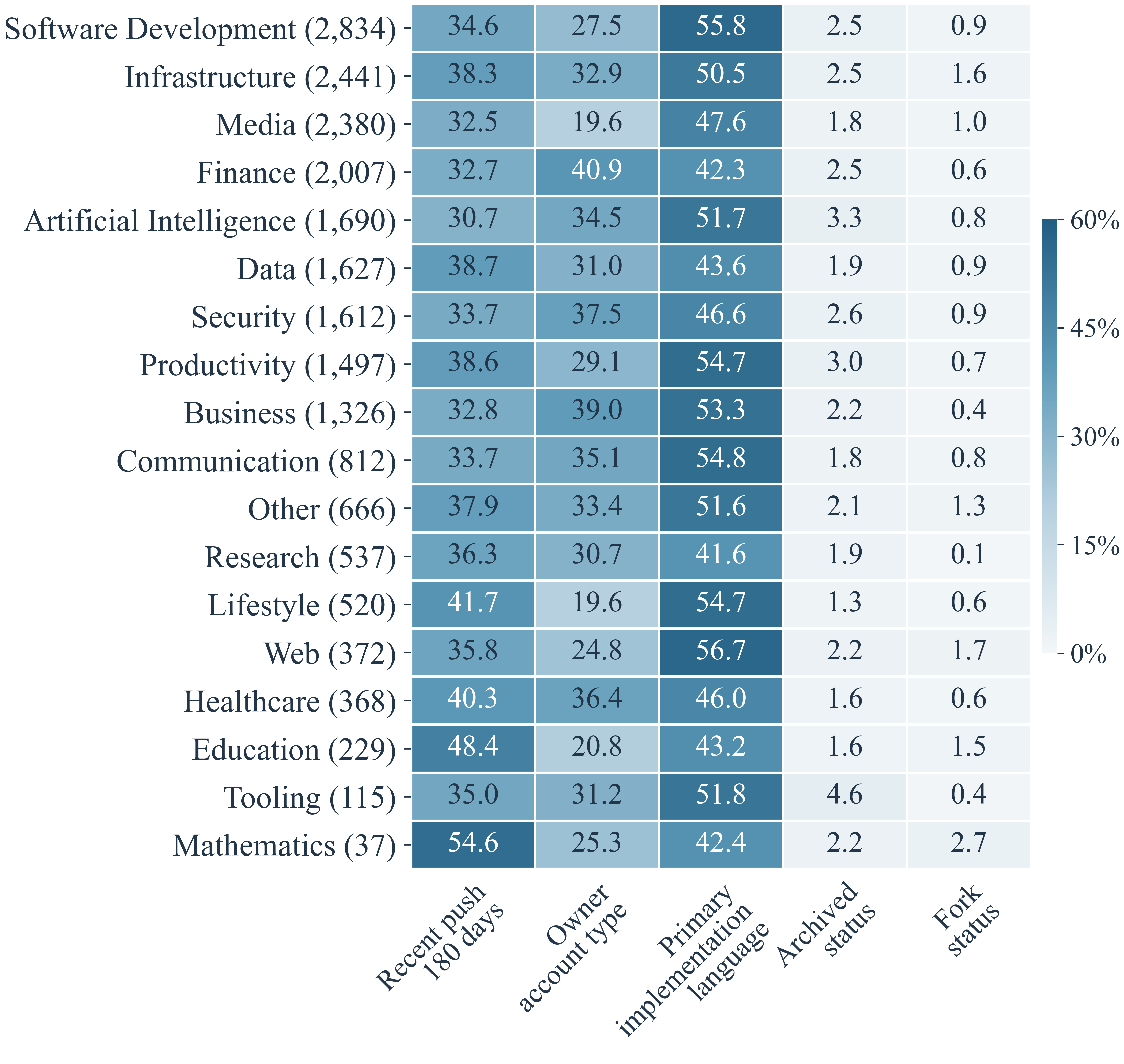}
    \caption{Within-group disagreement $\bar{D}_a$ (\%) among projects hosting functionally comparable tools.}
    \label{fig:fig04_domain_project_heterogeneity}
\end{wrapfigure}

\vspace{2pt}\noindent\textbf{Finding 8: Comparable tools come from very different projects: in 60.4\% of groups, some candidate projects are recently active while others are not.}

Beyond the code itself, users weigh project-level signals such as recent activity and ownership when choosing a tool; we therefore ask whether these signals also vary among functionally comparable alternatives.

We analyze the 21{,}070 groups whose tools come from at least two repositories with available metadata (97{,}279 tools from 29{,}772 repositories across all 18 domains), counting each repository once per group.
For five repository attributes, namely recent activity (not archived and pushed within 180 days of collection), owner type (user or organization), primary language, archived status, and fork status, we compute the share of repository pairs within a group that disagree, $D_{g,a} = \binom{m_{g,a}}{2}^{-1} \sum_{i<j} \mathbf{1}[x_{i,a} \neq x_{j,a}]$, where $m_{g,a} \ge 2$ is the number of repositories in group $g$ with an observed value $x_{i,a}$ of attribute $a$, and average it with equal weight across groups to obtain $\bar{D}_a$ (Figure~\ref{fig:fig04_domain_project_heterogeneity}).
Candidate projects differ most in primary language ($\bar{D} = 49.87\%$), recent activity (35.31\%), and owner type (31.36\%); in 60.38\% of groups, at least one candidate project is recently active and another is not.
Archived status (2.36\%) and fork status (0.91\%) rarely differ, likely because few repositories are archived or forked in the first place.
This heterogeneity is not confined to a few domains: across all 18 domains, disagreement ranges from 30.71\% to 54.60\% for recent activity and from 41.63\% to 56.73\% for primary language, while individual attributes vary independently, e.g., Media and Finance show nearly identical activity disagreement (32.50\% vs.\ 32.70\%) but very different owner-type disagreement (19.60\% vs.\ 40.90\%).
Functional matching thus leaves users with candidates of very different provenance and maintenance, so project metadata provides a further, independent basis for choosing among them, for example, to prefer an actively maintained alternative.

\answerbox{RQ3}{Functional comparability does not imply equivalence. Tools that perform the same operation can differ substantially in security alerts, code complexity and size, and project characteristics such as activity, ownership, and implementation language. In many cases, complexity or size differs by more than 2 times. The taxonomy identifies functionally comparable candidates, but selecting among them requires additional tool- and project-level signals.}

\subsection{RQ4: Downstream Task Utility.}

As the number of available MCP tools grows, an agent must identify the
appropriate set of tools to complete a user's task.
A flat candidate list exposes all tools at once, whereas \sys{} organizes tools by functional capability, allowing the agent to narrow the search space step by step.
RQ4 examines whether this functional taxonomy improves downstream task utility across models (Finding~9) and whether its benefits persist as the candidate set grows (Finding~10).

\vspace{2pt}\noindent\textbf{Evaluation Setup.}
We evaluate \sys{} on the same  100 released tasks from MCP-Atlas \cite{bandi2026mcp}, which contain multi-tool instructions and expert-annotated reference trajectories. We map all 220 unique benchmark tools to the \sys{} database and taxonomy. For each task, the flat list (baseline) and hierarchical taxonomy (\sys{}) receive exactly the same enabled tools and differ only in how those tools are organized and retrieved.

In the baseline, the model receives the complete flat list of enabled tools. In \sys{}, the same tools are organized into a task-specific subtree extracted from the global functional taxonomy. Starting at the root, the model retains clearly or potentially relevant branches and expands only
those branches. It then selects a complete tool plan from the tools reached through the retained paths. Thus, \sys{} uses the taxonomy for retrieval but does not change the tools available to the model.

We evaluate \texttt{GPT-5.6-Luna}, \texttt{GPT-5.6-Terra},
\texttt{GPT-5.5}, and \texttt{GPT-5.4} under both conditions, holding
task prompts and reasoning-effort settings fixed. Following MCP-Atlas,
we report two complementary metrics. \emph{Mean Claim Coverage} measures how fully task requirements are satisfied: each reference claim receives 1, 0.5, or 0 for full, partial, or no fulfillment, respectively, with scores averaged first within each task and then across tasks. \emph{Pass@0.75} measures the proportion of tasks achieving at least 75\% claim coverage. The former captures average coverage, including partial progress, while the latter captures how often a specified coverage target is reached. We use \texttt{GPT-5.4} to judge claim fulfillment with a fixed prompt and scoring rubric.


\begin{table*}[t]
\centering
\caption{
Performance comparison between the baseline (MCP-Atlas) and MCPacific on 100 tasks. The first section evaluates performance across models, while the second evaluates robustness of increasing candidate-tool set sizes with \texttt{GPT-5.6-Luna}.
$\Delta$ denotes MCPacific minus Baseline. 
}
\label{tab:model-comparison}
\resizebox{\textwidth}{!}{
\begin{tabular}{llccc ccc}
\toprule
\multirow{2}{*}{Evaluation} &
\multirow{2}{*}{Setting} &
\multicolumn{3}{c}{Mean Claim Coverage} &
\multicolumn{3}{c}{Pass@0.75} \\
\cmidrule(lr){3-5}
\cmidrule(lr){6-8}
&
&
Baseline &
MCPacific &
$\Delta$ &
Baseline &
MCPacific &
$\Delta$ \\
\midrule

\multirow{4}{*}{Model robustness}
& GPT-5.6-Luna
& 76.36\%
& \textbf{79.09\%}
& \textcolor{green!60!black}{$\uparrow$ 2.73\%}
& 61.00\%
& \textbf{70.00\%}
& \textcolor{green!60!black}{$\uparrow$ 9.00\%} \\

& GPT-5.6-Terra
& 75.22\%
& \textbf{78.71\%}
& \textcolor{green!60!black}{$\uparrow$ 3.48\%}
& 67.00\%
& \textbf{68.00\%}
& \textcolor{green!60!black}{$\uparrow$ 1.00\%} \\

& GPT-5.5
& 74.22\%
& \textbf{78.66\%}
& \textcolor{green!60!black}{$\uparrow$ 4.43\%}
& 66.00\%
& \textbf{68.00\%}
& \textcolor{green!60!black}{$\uparrow$ 2.00\%} \\

& GPT-5.4
& 73.65\%
& \textbf{74.33\%}
& \textcolor{green!60!black}{$\uparrow$ 0.69\%}
& 58.00\%
& \textbf{66.00\%}
& \textcolor{green!60!black}{$\uparrow$ 8.00\%} \\

\midrule

\multirow{3}{*}{Candidate-set robustness}
& Original
& 76.36\%
& \textbf{79.09\%}
& \textcolor{green!60!black}{$\uparrow$ 2.73\%}
& 61.00\%
& \textbf{70.00\%}
& \textcolor{green!60!black}{$\uparrow$ 9.00\%} \\

& Original $+30$
& \textbf{75.58\%}
& 74.90\%
& {$\downarrow$ 0.67\%}
& \textbf{64.00\%}
& \textbf{64.00\%}
& - \\

& Original $+50$
& 70.01\%
& \textbf{74.46\%}
& \textcolor{green!60!black}{$\uparrow$ 4.45\%}
& 54.00\%
& \textbf{66.00\%}
& \textcolor{green!60!black}{$\uparrow$ 12.00\%} \\

\bottomrule
\end{tabular}
}
\end{table*}




\vspace{2pt}\noindent\textbf{Finding 9: Taxonomy-guided retrieval improves task performance across all four models without adding tools.}
As shown in the first section of Table~\ref{tab:model-comparison}, \sys{} outperforms the flat-list baseline on both metrics for all four models. 
Mean Claim Coverage increases by 0.69 to 4.43 percentage points, while Pass@0.75 increases by 1 to 9 points. \texttt{GPT-5.5} obtains the largest coverage gain, increasing from 74.22\% to 78.66\%, whereas \texttt{GPT-5.6-Luna} obtains the largest Pass@0.75 gain, increasing from 61.00\% to 70.00\%.
The consistency of these improvements indicates that \sys{}'s benefit is not tied to a particular model. 
Organizing candidates by functional capability helps the model focus on task-relevant regions of the tool space and retrieve evidence supporting more reference claims.
A small average coverage gain can produce a large Pass@0.75 increase when several tasks move from slightly below to slightly above the 0.75 threshold.

\vspace{2pt}\noindent\textbf{Finding 10: \sys{} is more robust when the candidate-tool space becomes crowded.}
To test the efficiency to a broader tool scope, we use \texttt{GPT-5.6-Luna} and augment each task's original enabled-tool set with 30 and 50 additional random tools. 
The original candidate sets contain an average of 15.37 tools per task, with a minimum of 7 and a maximum of 30.
The second section of Table~\ref{tab:mcp-toolscape-comparison} shows that \sys{} improves Mean Claim Coverage by 2.73 points and Pass@0.75 by 9 points on the original candidate sets. 
With 30 additional tools, the two methods perform similarly.
However, with 50 additional tools, the difference becomes substantially larger. \sys{} improves
Mean Claim Coverage from 70.01\% to 74.46\% and Pass@0.75 from 54.00\% to 66.00\%, corresponding to gains of 4.45 and 12 points, respectively. 
These results indicate that hierarchical organization becomes more useful when the flat candidate list contains more tools and broader tool scope. The baseline must consider every tool in a single selection context, whereas \sys{} can discard unrelated functional branches before final tool selection. Its advantage is more obvious in the larger candidate-set condition, where flat-list retrieval experiences the greatest degradation.

\answerbox{RQ4}{\sys{} consistently improves task-level utility across models and is more robust to large candidate-tool sets than flat-list retrieval (baseline). Its advantage becomes more pronounced as the tool space grows, demonstrating that functional taxonomy effectively supports scalable tool retrieval and selection.}

\section{Threats to Validity}
\label{sec:threats}
We discuss threats to validity following Wohlin et al.~\cite{wohlin2012experimentation}.

\vspace{2pt}\noindent\textbf{Internal validity.}
The main internal threat is error in our automated pipeline.
Static extraction may miss dynamically registered tools or resolve entry functions incorrectly. We mitigate this by deriving 30 recognition patterns through manual inspection of implementations in all seven languages and excluding tools with missing or unreadable descriptions.
LLMs may also misclassify tools or behave non-deterministically. We mitigate this by evaluating each category design on a separate pilot sample, checking each node for mutually exclusive and exhaustive categories, sending unresolved cases to human review, and validating final assignments with an LLM judge calibrated against two human annotators (Section~\ref{sec:taxonomy-eval}).
Semgrep~\cite{semgrep} rules can produce false positives and negatives, and an LLM further maps alerts to categories. We therefore interpret these alerts as potential, rather than confirmed, issues and report the corresponding rule-level evidence.
Finally, in RQ4, \texttt{GPT-5.4} serves both as an evaluated model and as the judge. Because the same judge, prompt, and rubric are applied to both the baseline and \sys{}, our analysis focuses on their relative differences rather than absolute scores.

\vspace{2pt}\noindent\textbf{Construct validity.}
Our measures approximate the properties of interest.
CCN and NLOC are established proxies for control-flow complexity and implementation size~\citep{landman2016empirical}, but measure only a tool's entry function rather than its transitive callees.
GitHub stars and recent pushes are also imperfect proxies for popularity and maintenance~\cite{borges2018s} and reflect only the state at collection time.

\vspace{2pt}\noindent\textbf{External validity.}
Our dataset covers 17 marketplaces but includes only source-available servers written in the seven most common languages, which account for 88.9\% of collected repositories; closed-source and remote-only servers are outside our scope.
The taxonomy is initially constructed from 5{,}000 servers and may therefore miss rare capabilities, although full-corpus mapping, leaf re-expansion, and human refinement extend its coverage.
Because the MCP ecosystem evolves rapidly, our findings represent a snapshot at collection time. However, the automated collection, extraction, and mapping pipeline enables the analysis to be repeated on future snapshots.

\section{Conclusion}
Its central lesson is that the open tool ecosystem is best understood through what tools do, not where they are listed: viewed functionally, the ecosystem offers abundant but uneven alternatives whose differences matter, and this structure is directly usable by agents.
For platform and agent developers, this suggests organizing tools by function to make discovery and fallback tractable at scale; for users, it means that choosing among functionally comparable tools requires security, code, and maintenance signals that marketplaces do not yet provide.
In future work, we plan to track the ecosystem across snapshots, verify substitutability between comparable tools at the schema and behavior level, and integrate \sys{} into agent frameworks for open-world tool selection.





\section*{Ethics and Privacy Statement}
Our study analyzes publicly available MCP marketplaces, repository metadata, and tool specifications. We do not collect private user interactions, credentials, or other intentionally non-public data, and we report aggregate results rather than evaluating individuals. Nevertheless, consolidating tools into a searchable taxonomy may increase the discoverability of capabilities that could be misused, while automated classification may reproduce marketplace biases or incorrectly characterize particular tools. Thus, we expose category definitions and validation results, and treat assignments as analytical annotations rather than security or quality endorsements. We will release only the information necessary for reproducibility, and recommend that downstream users independently examine a tool's provenance, permissions, maintenance, and security before deployment.

\bibliographystyle{ACM-Reference-Format}
\bibliography{paper}

\appendix

\end{document}